\documentclass[aps,preprint,preprintnumbers,amsmath,amssymb,nofootinbib]{revtex4}
\usepackage{graphicx}
\usepackage{epsfig}
\newcommand {\beg}{\begin{equation}}
\newcommand {\en}{\end{equation}}
\newcommand {\bega}{\begin{eqnarray}}
\newcommand {\ena}{\end{eqnarray}}
\begin{document}
\title{Gravito-magnetic currents in the inflationary universe from WIMT}
\author{$^{1,2}$
Jes\'us Mart\'{\i}n Romero\footnote{E-mail address:
jesusromero@conicet.gov.ar}, $^{1,2}$ Mauricio Bellini
\footnote{E-mail address: mbellini@mdp.edu.ar} }
\address{$^1$ Departamento de F\'isica, Facultad de Ciencias Exactas y
Naturales, Universidad Nacional de Mar del Plata, Funes 3350, C.P.
7600, Mar del Plata, Argentina.\\
$^2$ Instituto de Investigaciones F\'{\i}sicas de Mar del Plata (IFIMAR), \\
Consejo Nacional de Investigaciones Cient\'ificas y T\'ecnicas
(CONICET), Argentina.}

\begin{abstract}
Using the Weitzenb\"ock representation of a Riemann-flat 5D
spacetime, we study the possible existence of primordial
gravito-magnetic currents from Gravito-electromagnetic Inflation (GEMI). We found that these currents
decrease exponentially in the Weitzenb\"ock representation, but
they are null in a Levi-Civita representation because we are
dealing with a 5D Riemann-flat spacetime without structure or
torsion.
\end{abstract}
\maketitle
\section{Introduction and Motivation}

It is well known that magnetic monopoles have been elusive as regards detection, despite the efforts. Such monopoles arise as a
theoretical possibility from the dual formulation of an
electrodynamic theory\cite{5}. The fate of primordial monopoles is
very closely linked to the history of the very early universe.
Preskill\cite{p} realized that this possible monopole production
could create a crisis for cosmology, implying far more monopoles
than observational limits allow. Because the expected energy scale
of grand unification is quite high, the geometrical size of a
monopole core must be quite small. Linde and Vilenkin
independently pointed out that such monopoles could expand
exponentially in the context of inflationary cosmology \cite{lv}.

However, there is an even more interesting possibility, which
arises from a theory that extends and unifies conceptually the
electrodynamics with a theory of gravity; gravito-electrodynamics. This theory was first outlined in
2006\cite{mb} in a cosmological context and later studied in
greater detail\cite{folcomp,membiela}, but the dual formalization
has not yet been addressed. In this paper we shall study the dual
formalism from a 5D vacuum and we will try and formalize their
dual streams, which, in a gravito-electrodynamic context are
related gravito-magnetic currents. This formalism is inspired by
the Induced Matter Theory (IMT), which is based on the assumption
that ordinary matter and physical fields that we can observe in
our 4D universe can be geometrically induced from a 5D Ricci-flat
metric with a space-like noncompact extra dimension on which we
define a physical vacuum\cite{IMT}. The Campbell-Magaard
theorem\cite{campbell,campbellb,campbellc,campbelld} serves as a
ladder to move between manifolds whose dimensionality differs by
one. This theorem, which is valid in any number of dimensions,
implies that every solution of the 4D Einstein equations with
arbitrary energy momentum tensor can be embedded, at least
locally, in a solution of the 5D Einstein field equations in
vacuum. Because of this, the stress-energy may be a 4D
manifestation of the embedding geometry. An extension of the IMT
was realized recently using the Weitzenb\"ock Induced Matter
Theory (WIMT)\cite{ruso}. This approach makes possible a
geometrical representation of a 5D vacuum (with a zero curvature
in the Weitzenb\"ock representation), on a nonzero curvature tensor
(in the sense of the Levi-Civita representation).

\section{Weitzenb\"{o}ck Induced Matter Theory (WIMT).}

We consider the basic elements for the extension of the IMT to a
geometrical description with the Weitzenb\"{o}ck connections. The
connections are constructed from certain 5D vielbeins related to
the transformation defined by
\begin{equation}\label{trafoconoco}
 \overrightarrow{e}_a = e^A_{\,\,a} \overrightarrow{E}_A,
\end{equation}
where $\overrightarrow{E}_A$ is an element of a base
$\{\overrightarrow{E}_A\}$ that we shall "initial base" (IB). In
our case we shall work with a 5D Minkowski space. Furthermore
$\overrightarrow{e}_a$ is an element of the "final base" (FB),
which is obtained trough the transformation (\ref{trafoconoco}).
In general, $\left\{\overrightarrow{E}_A\right\}$ can not be
coordinated\footnote{\begin{itemize} \item Capital latin letters
$A,B,C,..,H=0,1,2,3,4$ run on the 5D "initial space" (SS). \item
Lowercase latin letters $a,b,c,..,h=0,1,2,3,4$ run on the 5D
"final space" (AS). \item Greek letters
$\alpha,\beta,...=0,1,2,3$ run on the 4D hypersurface embedded in
the AS. \item Indices $i,j,k,...=1,2,3$ and $I,J,K,...=1,2,3$ run
on the 3D purely space of the AS and SS, respectively.
\end{itemize}}.

We can write the components of any tensor trough the 5D vielbein
$e^A_{\,\,a}$ and their inverses $\bar{e}^a_{\,\,A}$, which comply
with $e^A_{\,\,a} \bar{e}^b_{\,\,A} = \delta^b_a$ and with
$e^A_{\,\,a} \bar{e}^a_{\,\,B} = \delta^A_B$. In particular for
the metric tensor, we have
\begin{equation}
g_{ab} = e^A_{\,\,a}e^B_{\,\,b} g_{AB}.
\end{equation}
If we use the Weitzenb\"ock connections
$^{(W)}\Gamma^c_{ab}=\bar{e}^c_N\overrightarrow{e}_b(e_a^N)$, it can
be seen that
\begin{eqnarray}
^{(W)}\nabla_{\overrightarrow{e}_b}\left(\overrightarrow{E}_A\right)&=&
\,^{(W)}\nabla_{\overrightarrow{e}_b}\left(e^a_A\overrightarrow{e}_a\right)
=\underbrace{e^a_A\left\{^{(W)}\Gamma^c_{ab}-e^c_N\overrightarrow{e}_b
\left(e^N_a\right)\right\}}\overrightarrow{e}_c=0.\nonumber\\
&\,&\,\,\,\,\,\,\,\,\,\,\,\,\,\,\,\,\,\,\,\,\,\,\,\,\,\,\,\,\,\,\,\,\,\,\,\,\,\,\,\,\,\,\,\,\,\,\,\,\,\,\,\,\,\,\,\,\,\,\,\,\,\,\,\,\,\,[^{(W)}\nabla_{
\overrightarrow{e}_b}(\overrightarrow{E}_A)]^c=
{[\overrightarrow{E}_A]^c}_{;\,b}=\bar{e}^c_{A;\,b}\label{dervielb}
\end{eqnarray}
It can be seen that the expressions
$[^{(W)}\nabla_{\overrightarrow{e}_b}(\overrightarrow{E}_A)]^c=
{[\overrightarrow{E}_A]^c}_{;\,b}$ represent the $c$-component of
the application of the derivative operator characterized by the
Weitzenb\"ock connections with respect to the $b$-component of the
base of the FB over the $A$-component of the vector of the ST. In
general, for any vector field
$\overrightarrow{A}=A^c\overrightarrow{e}_c$,
\begin{footnote}{We denote by ";" the covariant derivative  \begin{eqnarray}\nonumber
\overrightarrow{A}&=&A^b\overrightarrow{e}_b=A^0\overrightarrow{e}_0+A^i\overrightarrow{e}_i
\Rightarrow \\\nonumber
\nabla_{\overrightarrow{e}_b}(\overrightarrow{A})&=&\nabla_{\overrightarrow{e}_b}(A^0
\overrightarrow{e}_0)+\nabla_{\overrightarrow{e}_b}(A^i\overrightarrow{e}_i)=\overrightarrow{e}_b(A^0)\overrightarrow{e}_0
+A^0\underbrace{\nabla_{\overrightarrow{e}_b}(\overrightarrow{e}_0)}+\overrightarrow{e}_b(A^i)\overrightarrow{e}_i+A^i
\Gamma^c_{ib}\overrightarrow{e}_c
=\\\nonumber &\,&\,\,\,\,\,\,\,\,\,\,\,\,\,\,\,\,\,\,\,\,\,\,\,\,\,\,\,\,\,\,\,\,\,\,\,\,\,\,\,\,\,\,\,\,\,\,\,\,\,\,\,\,\,\,\,\,\,\,\,\,\,\,\,\,\,\,\,\,\,\,\,\,\,\,\,\,\,\,\,\,\,\,\,\,\,\,\,\,\,\,\,\,\,\,\,\,\,\,\,\,\,\,\,\,\,\,\,\,\,\,\,\,\,\,\,\,\,\,\,\,\,\,\,\,\,\,\,\,\,\,\,\,\, \Gamma^c_{0b}\overrightarrow{e}_c\\
\nonumber
&=&\left(\overrightarrow{e}_b(A^0)+A^0\Gamma^0_{0b}+A^i\Gamma^0_{ib}\right)\overrightarrow{e}_0+\left(
\overrightarrow{e}_b(A^i)+A^0\Gamma^i_{0b}+A^j\Gamma^i_{jb}\right)\overrightarrow{e}_i.\end{eqnarray}
Therefore
$[\nabla_{\overrightarrow{e}_b}(\overrightarrow{A})]^0=A^0_{\,;\,b}$.
This is not only valid for the superscript $0$, but also for
the other indices, so that
$[\nabla_{\overrightarrow{e}_b}(\overrightarrow{A})]^c=A^c_{\,;\,b}$.}\end{footnote}
we have ${[\overrightarrow{A}]^c}_{;\,b}= A^c_{;\,b}$.

When the field $\overrightarrow{A}$ is the $A$-component of the
IB, $\overrightarrow{A}=\overrightarrow{E}_A$, the vector field
is given by
$\overrightarrow{E}_A=\bar{e}^a_A\overrightarrow{e}_a$ and it is
clear that $[\overrightarrow{E}_A]^c=\bar{e}^c_A$. In this sense
the connections of Weitzenb\"{o}ck imply $\bar{e}^c_{A;\,b}=0$,
and the vielbeins are seen as coefficients of the development of
one base in another base. The expression (\ref{dervielb}) is
fulfilled if we are working with the Weitzenb\"{o}ck connections.

Now we consider a 5D spacetime described by the metric $g_{AB}$ in
the IS, and $g_{ab}$ describing the FS. It is obvious that the latter
space is Weitzenb\"ock-flat in the sense that the Riemann
tensor constructed through this kind of connections is null:
$^{(W)}R^a_{bcd}=0$. However, it cannot be Riemann-flat with
respect to the Levi-Civita connections: $^{(lc)}{R}^a_{bcd}\neq
0$. The Riemann tensor written with the Weitzenb\"ock
representation for the spacetime characterized by the metric
$g_{ab}$, is given by
{\footnotesize{\begin{eqnarray}\label{riew}\\\nonumber
^{(W)}R^a_{bcd}
&=&\,\overrightarrow{e}_b\left(^{(W)}\Gamma^a_{dc}\right) -
\overrightarrow{e}_c\left(^{(W)}\Gamma^a_{db}\right) +
\,^{(W)}\Gamma^n_{dc} \,^{(W)}\Gamma^a_{nb} -
^{(W)}\Gamma^n_{db}\,^{(W)}\Gamma^a_{nc}- C^n_{cb}\,
^{(W)}\Gamma^a_{dn}=0, \label{r1}
\end{eqnarray}}}
where $^{(W)}\Gamma^a_{bc}$ are the Weitzenb\"ock connections and
${C}^a_{bc}$ are the coefficients of structure of the FB, which
can be expressed through
$C^a_{bc}=\bar{e}^a_N\overrightarrow{e_c}(e^N_b)-\bar{e}^a_N\overrightarrow{e_b}(e^N_c)=\,^{(W)}\Gamma^a_{bc}-\,^{(W)}\Gamma^a_{cb}$,
when the absence of structure of the IB makes the
Weitzenb\"ock torsion mull. Both representations are related by the
expression
\begin{equation}\label{relacionwlc}
^{(W)}\Gamma^a_{bc} = ^{(lc)}{\Gamma}^a_{bc} - \,^{(W)}K^a_{bc},
\end{equation}
where the Weitzenb\"ock contortion $^{(W)}K^a_{bc}$ is related to
the Weitzenb\"ock torsion $^{(W)}T^a_{bc}$
\begin{equation}\label{contorsiong}^{(W)}K^a_{bc}=\frac{g^{ma}}{2}\{g_{bm;c}
+g_{mc;b}-g_{bc;m}\}+\frac{g^{ma}}{2}\{^{(W)}T^n_{cm}g_{bn}+\,^{(W)}T^n_{bm}g_{nc}-\,^{(W)}T^n_{cb}g_{nm}\}.\end{equation}
Here, we have considered the nonzero non-metricity
$g_{ab;\,c}=\,^{(W)}Q_{abc}$. When $g_{ab;\,c}=0$, the tensor
$^{(W)}K^a_{bc}$ reduces to the well-known contortion tensor in
the Weitzenb\"ock representation
$^{(W)}K^a_{bc}=\frac{g^{ma}}{2}\{^{(W)}T^n_{cm}g_{bn}+\,^{(W)}T^n_{bm}g_{nc}-\,^{(W)}T^n_{cb}g_{nm}\}$.
\begin{footnote}{In this work we shall use this definition with zero nonmetricity.}\end{footnote}.

On the other hand, if the Weitzenb\"{o}ck torsion becomes zero (this
holds when the IB has no structure), we have
$^{(W)}K^a_{bc}=\frac{g^{ma}}{2}\{g_{bm;c}+g_{mc;b}-g_{bc;m}\}$.

By contracting the null-tensor $^{(W)}R^a_{bcd}$ we obtain the
following tensors: $^{(W)}S_{bc}=\,^{(W)}R^a_{bca}$ (which is
antisymmetric), and $^{(W)}R_{cd}=\,^{(W)}R^a_{acd}$ (which is
symmetric), that is\footnote{The expressions
$^{(W)}\Gamma^n_{(d|a}\,^{(W)}\Gamma^a_{n|c)}$ indicate the
symmetrization of the indices $d$ and $c$, inside the parentheses,
but excepting the indices $a$ and $n$ inside the vertical bars
$"|"$. }
{{\begin{eqnarray}\label{r1}&\,&\\\nonumber
^{(W)}S_{bc}=\,^{(W)}R^a_{bca} &=
&\,\overrightarrow{e}_b\left(^{(W)}\Gamma^a_{ac}\right) -
\overrightarrow{e}_c\left(^{(W)}\Gamma^a_{ab}\right) +
\,^{(W)}\Gamma^n_{ac} \,^{(W)}\Gamma^a_{nb}\\\nonumber &\,& -
^{(W)}\Gamma^n_{ab}\,^{(W)}\Gamma^a_{nc}- C^n_{cb}\,
^{(W)}\Gamma^a_{an}=0, \\\label{r2}&\,&\\\nonumber
^{(W)}R_{cd}=\,^{(W)}R^a_{a(cd)} &=
&\,\overrightarrow{e}_a\left(^{(W)}\Gamma^a_{(dc)}\right) -
\overrightarrow{e}_{(c}\left(^{(W)}\Gamma^a_{d)a}\right) +
\,^{(W)}\Gamma^n_{(dc)} \,^{(W)}\Gamma^a_{na}\\\nonumber &\,& -
^{(W)}\Gamma^n_{(d|a}\,^{(W)}\Gamma^a_{n|c)}- C^n_{(c|a}\,
^{(W)}\Gamma^a_{|d)n}=0.
\end{eqnarray}}}

From (\ref{riew}), (\ref{relacionwlc}), (\ref{r1}) and (\ref{r2})
we obtain the expressions for the corresponding curvature
tensors with the Weitzenb\"{o}ck representation by means of those
of the Levi-Civita representation (and vice versa). Hence, we have
\begin{eqnarray}\label{riewlc}^{(W)}R^a_{bcd}&=&\,^{(lc)}R^a_{bcd}-\overrightarrow{e}_b(^{(W)}K^a_{dc})
+\overrightarrow{e}_c(^{(W)}K^a_{db})\\\nonumber
&\,&-\,^{(W)}K^n_{dc}\,^{(W)} K^a_{nb} -\,^{(lc)}\Gamma^n_{dc}
\,^{(W)} K^a_{nb} -
\,^{(W)}K^n_{dc}\,^{(lc)}\Gamma^a_{nb}\\\nonumber
&\,&+\,^{(W)}K^n_{db}\,^{(W)} K^a_{nc}+\,^{(lc)} \Gamma^n_{db}
\,^{(W)}K^a_{nc}+\,^{(W)}K^n_{db}\,^{(lc)}\Gamma^a_{nc}\\\nonumber
&\,&+C^n_{cb}\,^{(W)}K^a_{dn}.\end{eqnarray}

Using the eqs. (\ref{r1}) and (\ref{r2}) in the last expression we
found the analogous equations for $^{(W)}S_{bc}$ and
$^{(W)}R_{cd}$, or for $^{(lc)}S_{bc}$ and $^{(lc)}R_{cd}$.
Usually it is simple to determinate the latter, because they comply
with the transformation $^{(lc)}S_{bc}=e^B_b
e^C_c\,^{(lc)}S_{BC}=0$ and $^{(lc)}R_{cd}=e^D_d
e^C_c\,^{(lc)}R_{CD}$, where, if the IB has no structure, the
tensors with capital indices are calculated in a coordinate base
with the Levi-Civita connections.

Now we shall consider the Einstein equations with the
Weitzenb\"ock representation. We shall try to obtain the effective
4D equations after making a constant foliation from a 5D
Weitzenb\"ock vacuum. Taking (\ref{r2}), we obtain
that {\scriptsize{\begin{eqnarray}
\left.\overbrace{^{(W)}{R}_{\zeta\delta}}^{5D}\,\right|_{l=l_0} &=
&\left.\overbrace{\overrightarrow{e}_{\alpha}\left(^{(W)}{\Gamma}^{\alpha}_{(\zeta\delta)}\right)
-
\overrightarrow{e}_{(\zeta}\left(\,^{(W)}{\Gamma}^{\alpha}_{\delta)\alpha}\right)
+ ^{(W)}{\Gamma}^\nu_{(\zeta\delta)}\,
^{(W)}{\Gamma}^{\alpha}_{\nu\alpha} -
^{(W)}{\Gamma}^{\epsilon}_{(\delta|\alpha}
 \,^{(W)}{\Gamma}^{\alpha}_{\nu|\zeta)}
- C^{\nu}_{(\zeta|\alpha}\,^{(W)}{\Gamma}^{\alpha}_{|\delta)\nu}}^{5D}\,\right|_{l=l_0} \nonumber \\
&\,&+ \left[
\left(\overrightarrow{e}_4\left(^{(W)}{\Gamma}^4_{(\zeta\delta)}\right)
-
\overrightarrow{e}_{(\zeta}\left(^{(W)}{\Gamma}^{4}_{\delta)4}\right)\right) \right. \nonumber \\
&\,&+ \left.\left( ^{(W)}{\Gamma}^4_{(\zeta\delta)}\,
^{(W)}{\Gamma}^{\alpha}_{4\alpha} +\,
^{(W)}{\Gamma}^{\nu}_{(\zeta\delta)}\, ^{(W)}{\Gamma}^{4}_{\nu 4}
+ \,^{(W)}{\Gamma}^{4}_{(\zeta\delta)}
\,^{(W)}{\Gamma}^4_{44} \right.\right. \nonumber \\
&\,&- \left.\left.\left.\,^{(W)}{\Gamma}^{4}_{(\delta|\alpha}
\,^{(W)}{\Gamma}^{\alpha}_{4|\zeta)} -
\,^{(W)}{\Gamma}^{\nu}_{(\delta|4}
\,^{(W)}{\Gamma}^{4}_{\nu|\zeta)} -
\,^{(W)}{\Gamma}^{4}_{(\delta|4}\,^{(W)}
{\Gamma}^4_{4|\zeta)}\right)\right.\right. \nonumber
\\
&\,&- \left.\left.\left(C^{4}_{(\zeta|\alpha}
\,^{(W)}{\Gamma}^{\alpha}_{|\delta)4} +
C^{4}_{(\zeta|4}\,^{(W)}{\Gamma}^4_{|\delta)\nu}
+C^{4}_{(\zeta|4}\,^{(W)}{\Gamma}^{4}_{|\delta)4}\right)
\right]\right|_{l=l_0}=0. \label{ind}
\end{eqnarray}}}

In this work we shall deal with canonical metrics\cite{..}. An
interesting example is\cite{LB}
\begin{equation}
dS^2= \left(\frac{l}{l_0}\right)^2
h_{\alpha\beta}(y^{\gamma})\,dy^{\alpha} dy^{\beta}-dl^2,
\end{equation}
where $l$ is related to the noncompact extra dimension and $l_0$
is a constant. After making the constant foliation, we obtain
$\overbrace{^{(W)}{\Gamma}^\epsilon_{\beta\alpha}}^{5D}|_{l=l_0}
=\overbrace{^{(W)}{\Gamma}^\epsilon_{\beta\alpha}}^{4D}$, where
$\overbrace{^{(W)}{\Gamma}^\epsilon_{\beta\alpha}}^{4D}$ is a
Weitzenb\"ock connection defined on the embedded 4D hypersurface
obtained through the foliation: $l=l_0$. It makes it possible to
obtain the effective 4D Ricci-Weitzenb\"{o}ck tensor
\begin{eqnarray}\label{ricci4w}\overbrace{^{(W)}{R}_{\zeta\delta}}^{4D}&=&- \left[
\left(\overrightarrow{e}_4\left(^{(W)}{\Gamma}^4_{(\zeta\delta)}\right)
-
\overrightarrow{e}_{(\zeta}\left(^{(W)}{\Gamma}^{4}_{\delta)4}\right)\right) \right. \\
&\,&+ \left.\left( ^{(W)}{\Gamma}^4_{(\zeta\delta)}\,
^{(W)}{\Gamma}^{\alpha}_{4\alpha} +\,
^{(W)}{\Gamma}^{\nu}_{(\zeta\delta)}\, ^{(W)}{\Gamma}^{4}_{\nu 4}
+ \,^{(W)}{\Gamma}^{4}_{(\zeta\delta)}
\,^{(W)}{\Gamma}^4_{44} \right.\right. \nonumber \\
&\,&- \left.\left.\left.\,^{(W)}{\Gamma}^{4}_{(\delta|\alpha}
\,^{(W)}{\Gamma}^{\alpha}_{4|\zeta)} -
\,^{(W)}{\Gamma}^{\nu}_{(\delta|4}
\,^{(W)}{\Gamma}^{4}_{\nu|\zeta)} -
\,^{(W)}{\Gamma}^{4}_{(\delta|4}\,^{(W)}
{\Gamma}^4_{4|\zeta)}\right)\right.\right. \nonumber
\\
&\,&- \left.\left.\left(C^{4}_{(\zeta|\alpha}
\,^{(W)}{\Gamma}^{\alpha}_{|\delta)4} +
C^{4}_{(\zeta|4}\,^{(W)}{\Gamma}^4_{|\delta)\nu}
+C^{4}_{(\zeta|4}\,^{(W)}{\Gamma}^{4}_{|\delta)4}\right)
\right]\right|_{l=l_0} \label{ind},\nonumber \end{eqnarray} which
is symmetric with respect to the indices $\zeta,\delta$. The
antisymmetric tensor is obtained as
$\overbrace{^{(W)}{S}_{\beta\zeta}}^{5D}|_{l=l_0}=
\overbrace{\,^{(W)}R^{a}_{\beta\zeta a}}^{5D}|_{l=l_0}=0$, so that
we obtain
\begin{eqnarray}\label{ricci4w}\overbrace{^{(W)}{S}_{\beta\zeta}}^{4D}&=&- \left[
\left(\overrightarrow{e}_{\beta}\left(^{(W)}{\Gamma}^4_{4\zeta}\right)
-
\overrightarrow{e}_{\zeta}\left(^{(W)}{\Gamma}^{4}_{4\beta}\right)\right) \right. \\
&\,&+ \left.\left( ^{(W)}{\Gamma}^4_{\alpha\zeta}\,
^{(W)}{\Gamma}^{\alpha}_{4\beta} +\,
^{(W)}{\Gamma}^{\nu}_{4\zeta}\, ^{(W)}{\Gamma}^{4}_{\nu \beta}\right.\right. \nonumber \\
&\,&- \left.\left.\left.\,^{(W)}{\Gamma}^{4}_{\alpha\beta}
\,^{(W)}{\Gamma}^{\alpha}_{4\zeta} -
\,^{(W)}{\Gamma}^{\nu}_{4\beta} \,^{(W)}{\Gamma}^{4}_{\nu\zeta}
\right)\right.\right. \nonumber
\\
&\,&- \left.\left.\left(C^{4}_{\zeta\beta}
\,^{(W)}{\Gamma}^{\alpha}_{\alpha4} +
C^{\nu}_{\zeta\beta}\,^{(W)}{\Gamma}^4_{4\nu}
+C^{4}_{\zeta\beta}\,^{(W)}{\Gamma}^{4}_{44}\right)
\right]\right|_{l=l_0} \label{ind}.\nonumber \end{eqnarray} The
Ricci-Weitzenb\"ock scalar curvature can be obtained from a 5D
vacuum, as
\begin{equation}\label{r4}
\overbrace{^{(W)}{R}}^{5D} =
\overbrace{g^{ab}\,^{(W)}{R}_{ab}}^{5D} =
\overbrace{g^{\alpha\beta}\,^{(W)}{R}_{\alpha\beta}}^{5D} +
\overbrace{g^{55}\,^{(W)}{R}_{55}}^{5D}=0.
\end{equation}
Hence, from eq. (\ref{r4}) we can obtain
\begin{equation}\overbrace{^{(W)}{ R}}^{4D} =
\left.\,\overbrace{g^{\beta\gamma}\,^{(W)}{R}_{\beta\gamma}}^{5D}\,\right|_{l=l_0}
=\overbrace{h^{\beta\gamma}\,^{(W)}{R}_{\beta\gamma}}^{4D},\end{equation}
which means that the scalar Ricci-Weitzenb\"ock curvature has the
source
\begin{equation}\overbrace{^{(W)}{R}}^{4D}=-\left.
\overbrace{g^{44}\,^{(W)}{R}_{44}}^{5D}\,\right|_{l=l_0},\end{equation}
and finally the induced Einstein-Cartan-Weitzenb\"ock equations
are
\begin{eqnarray}
\overbrace{^{(W)}{G}_{\beta\gamma}}^{4D}& = &
\overbrace{^{(W)}{R}_{\beta\gamma}}^{4D} - \frac{1}{2}
 \overbrace{h_{\beta\gamma}\,^{(W)}{ R}}^{4D} = -8\pi G \,
\overbrace{\bar{T}_{(\beta\gamma)}}^{4D}, \label{EW}
\\
\overbrace{^{(W)}{S}_{\beta\gamma}}^{4D} &=& - 8\pi G\overbrace{\bar{T}_{[\beta\gamma]}}^{4D}, \label{EW2} \\
^{(W)}{R}_{a 5} & = & 0, \label{ew1} \\
^{(W)}{S}_{a 5} & = & 0, \label{ew2}
\end{eqnarray}
where we have taken into account in (\ref{EW2}) that
$g^{\beta\gamma}\,^{(W)}{S}_{\beta\gamma}=0$. Furthermore, the symmetric and antisymmetric
parts of the energy-momentum tensor in eqs. (\ref{EW}) and (\ref{EW2}) are given by $\overbrace{\bar{T}_{(\beta\gamma)}}^{4D}={1\over
2}(\overbrace{\bar{T}_{\beta\gamma}}^{4D}+\overbrace{\bar{T}_{\gamma\beta}}^{4D})$
and $\overbrace{\bar{T}_{[\beta\gamma]}}^{4D}={1\over 2}
(\overbrace{\bar{T}_{\beta\gamma}}^{4D}-\overbrace{\bar{T}_{\gamma\beta}}^{4D})$.
The Equation (\ref{EW}) describe the dynamics of the gravitational field using the Weitzenb\"ock representation
on the effective 4D hypersurface .

\subsection{Effective 4D dynamics with the Levi-Civita representation.}

Now we shall intend to write the curvature and the Ricci tensors (in the Levi-Civita representation) with respect to the Weitzenb\"ock connections
and contortions. The Ricci tensor in the Levi-Civita representation is related to the Ricci tensor in the Weitzenb\"ock representation plus additional
terms that depend on contortions and structure
{\scriptsize{\begin{eqnarray}\label{lcSxx}\nonumber\overbrace{{^{(lc)}R}_{\zeta\delta}}^{4D}&=&
\left.\overrightarrow{e}_{\alpha}\left(^{(lc)}{\Gamma}^{\alpha}_{(\zeta\delta)}\right)
-
\overrightarrow{e}_{(\zeta}\left(\,^{(lc)}{\Gamma}^{\alpha}_{\delta)\alpha}\right)
+ ^{(lc)}{\Gamma}^\nu_{(\zeta\delta)}\,
^{(lc)}{\Gamma}^{\alpha}_{\nu\alpha} -
^{(lc)}{\Gamma}^{\epsilon}_{(\delta|\alpha}
 \,^{(lc)}{\Gamma}^{\alpha}_{\nu|\zeta)}
-
C^{\nu}_{(\zeta|\alpha}\,^{(lc)}{\Gamma}^{\alpha}_{|\delta)\nu}\,\right|_{l=l_0}\\
\nonumber &=&
^{(W)}R_{\zeta\delta}+\overrightarrow{e}_{\alpha}\left(K^{\alpha}_{(\zeta\delta)}\right)-
\overrightarrow{e}_{(\zeta}\left(K^{\alpha}_{{\delta)\alpha}}\right)+K^{\nu}_{(\zeta\delta)}
K^{\alpha}_{\nu\alpha}+\,^{(W)}\Gamma^{\nu}_{(\zeta\delta)}K^{\alpha}_{\nu\alpha}+K^{\nu}_{(\zeta\delta)}\,^{(W)}\Gamma^{\alpha}_{\nu\alpha}\\
&\,&-K^{\nu}_{(\delta|\alpha}K^{\alpha}_{\nu|\zeta)}-\,^{(W)}\Gamma^{\nu}_{(\delta|\alpha}
K^{\alpha}_{\nu|\zeta)}-K^{\nu}_{(\delta|\alpha}\,^{(W)}\Gamma^{\alpha}_{\nu|\zeta)}
-C^{\nu}_{(\zeta|\alpha}K^{\alpha}_{|\delta)\nu}.\end{eqnarray}}}
The scalar curvature is
\begin{equation}\label{lcResc}\overbrace{^{(lc)}{R}}^{4D}=-\left.
\overbrace{g^{44}\,^{(lc)}{R}_{44}}^{5D}\,\right|_{l=l_0}.\end{equation}
Hence, the Einstein-Cartan equations are given by
\begin{eqnarray}
\overbrace{^{(lc)}{G}_{\beta\gamma}}^{4D}& = &
\overbrace{^{(lc)}{R}_{\beta\gamma}}^{4D} - \frac{1}{2}
\overbrace{{h_{\beta\gamma}\, ^{(lc)}R}}^{4D} =  -8\pi G \,
\overbrace{{T}_{(\beta\gamma)}}^{4D}, \label{EWW}
\\
\overbrace{{S}_{\beta\gamma}}^{4D}& =  &-8 \pi G
\,\overbrace{{T}_{[\beta\gamma]}}^{4D}=0, \label{EW1}
\end{eqnarray}
joined with $^{(W)}{R}_{a5}=0$ and $^{(W)}{S}_{a5}=0$ that are
additional conditions. Here,
$\overbrace{{T}_{(\beta\gamma)}}^{4D}$ and
$\overbrace{{T}_{[\beta\gamma]}}^{4D}$ are the symmetric and
antisymmetric energy-momentum tensors induced on the 4D
hypersurface written in the Levi-Civita
representation\footnote{The equations (\ref{EW1}) take into
account the Cartan equations which describe spinor contributions.
\begin{eqnarray} \overbrace{{S}_{\beta\gamma}}^{4D} -
\frac{1}{2} \sigma_{\beta\gamma} S =-8 \pi G
\,\underbrace{\overbrace{{T}_{[\beta\gamma]}}^{4D}}_{spin},
\end{eqnarray}
where $S= \sigma^{\mu\nu} S_{\mu\nu}$, $\sigma^{\mu\nu}={1\over 2}
\left[\gamma^{\mu}, \gamma^{\nu}\right]$ and $\gamma^{\mu}$ are the
Dirac matrices.}.

\section{Dual action and equations of motion}

We shall consider the conditions by which we can induce curvature
and currents by means of WIMT\cite{ruso}, on a 5D spacetime
represented by cartesian coordinates. The 5D tensor metric can be
written as
\begin{equation}\label{metri}
[\eta]_{AB}= {\rm diag}\left[  1,-1,-1,-1,-1\right].
\end{equation}
We can construct a FB with interesting cosmological properties if
we take the appropriate vielbein. The action for the
gravito-electromagnetic fields in a 5D vacuum can be written in
the form
\begin{eqnarray}\label{accionfaraday}\mathcal{S}=\int d^5x\sqrt{|\eta|}\left[\frac{R}{16\,\pi\,G}-\frac{1}{4}
F_{AB}
F^{AB}-\frac{\lambda}{2}\left(A^B_{;\,B}\right)^2\right].\end{eqnarray}
In order to may be render the dual currents interesting, 
rewrite the last action in terms of the dual tensors ${\cal F}_{ABC}$
\begin{eqnarray}\label{act0}\\\nonumber
    \mathcal{S}_1 &=&\int d^5x\sqrt{|\eta|}\left[\frac{R}{16\,\pi\,G}-\frac{k}{4} {\cal F}_{ABC} {\cal F}^{ABC}-\frac{\lambda}{2}
    \left(A^B_{;\,B}\right)^2\right],\\\nonumber \mathcal{S}&=&\int d^5x\sqrt{|\eta|}\left[\frac{R}{16\,\pi\,G}-\frac{1}{4} F_{AB} F^{AB}
    -\frac{\lambda}{2}\left(A^B_{;\,B}\right)^2\right].\end{eqnarray}It is evident that
$\frac{1}{3!}\mathcal{F}_{ABC}\mathcal{F}^{ABC}
=\frac{1}{3!\,4}\varepsilon_{ABCDE}\varepsilon^{ABCNM}F^{DE}F_{NM}=F^{NM}F_{NM}$,
where we have used $\varepsilon_{ABCDE}
\varepsilon^{ABCNM}=3!\,2!\,(\delta^N_D\delta^M_E-\delta^N_E\delta^M_D)$, so that when $k=\frac{1}{3!}$, we see that
both actions describe the same physical system:
\begin{equation}\nonumber
\mathcal{S}_1 = \mathcal{S}.
 \end{equation}
In our case, when we use the Lorentz gauge and we deal with a 5D vacuum with $R=0$, we have $\mathcal{S}_1 \sim \mathcal{S}$
and both actions give us the same equations of motion. In order to describe the dual sources of these equations we shall deal with the action $\mathcal{S}_1$.
The dynamics of the gravito-electromagnetic fields obtained taking the extreme with the action (\ref{accionfaraday})), is
${A^K_{\,;\,B}}^{;B}-(1-\lambda){A^B_{\,;\,B}}^{;\,K}=0$. On a 5D vacuum  ($R=0$), and when we take $\lambda=1$ and the Lorentz gauge
$A^B_{;\,B}=0$, they reduce to
\begin{eqnarray}\label{dina0}
    \Box A^K&=&\eta^{BC} A^K_{;\,BC}=0,
\end{eqnarray}
which are the Klein-Gordon equations for massless fields. The gravito-magnetic currents come from
the solutions for the fields (\ref{dina0}). The last equations are compatible with a current which
has its source in
\begin{eqnarray}\label{6.00-1}F^{NB}_{;\,B}=-\eta^{AN}\left[A^M\,R^B_{ABM}+A^B_{;\,M}T^M_{BA}\right],
\end{eqnarray}
where we have used (\ref{dina0}) and the Lorentz gauge in absence
of nonmetricity. We have used the fact that
\begin{eqnarray}\nonumber
F^{NB}_{;\,B}&=&\eta^{AN}\eta^{BM}F_{AM;B}=g^{AN}A^B_{;\,AB}-
\underbrace{\eta^{BM}A^N_{;MB}},\\\nonumber
&\,&\,\,\,\,\,\,\,\,\,\,\,\,\,\,\,\,\,\,\,\,\,\,\,\,\,\,\,\,\,
\,\,\,\,\,\,\,\,\,\,\,\,\,\,\,\,\,\,\,\,\,\,\,\,\,\,\,\,\,\,\,\,\,\,\,\,\,\,\,\,\,\,\,\,\,\,\,\,\,\,\,\,
\,\,\,\,\,\,\,\,\,\,\,\,\,\,0\end{eqnarray} where clearly we have
imposed the absence of currents on the 5D vacuum with respect to
the Levi-Civita representation. Hence
{\footnotesize{\begin{eqnarray}\label{6.00}
^{(lc)}F^{NB}_{;\,B}&=&0\,\Rightarrow \\
\label{6.01}^{(W)}F^{NB}_{;B}&=&-\left(^{(W)}F^{MB}+
\eta^{RM}A^PK^B_{PR}-\eta^{RB}A^PK^M_{PR}\right)K^N_{MB}\\
\nonumber &\,&-\left(^{(W)}F^{NM}+
\eta^{RN}A^PK^M_{PR}-\eta^{RM}A^PK^N_{PR}\right)K^B_{MB},\end{eqnarray}}}
where $^{(lc)}F_{AB}=
\overrightarrow{E}_A(A_B)-\overrightarrow{E}_B(A_A)+A_N(C^N_{AB})$
y $^{(W)}F_{AB}=
\overrightarrow{E}_A(A_B)-\overrightarrow{E}_B(A_A)+A_N(^{(W)}T^N_{BA}+C^N_{AB})$
\begin{footnote}{Using the expression $^{(lc)}\Gamma^A_{BC}=^{(W)}\Gamma^A_{BC}
+K^A_{BC}$ it is possible to obtain the following expression
for both Faraday tensors:
$^{(lc)}F^{NB}=\,^{(W)}F^{NB}+\eta^{RN}A^PK^B_{PR}-\eta^{RB}A^PK^N_{PR}$.
If we make the derivative of this expression and we use
(\ref{6.00}), we can check the validity of eq.
(\ref{6.01}).}\end{footnote}.

Notice that in (\ref{6.00}) we use the covariant derivative
with respect to the Christoffel symbols, but in (\ref{6.01}) we use the derivative covariant with respect to the
Weitzenb\"{o}ck connections. Hence, we can adopt both
representations in a complementary mode to describe the 5D vacuum.

The Weitzenb\"{o}ck currents are related by one expression
analogous to eq. (\ref{6.00}), but it is more complicated
\begin{eqnarray}\label{6.03}^{(lc)(m)}J_{AB}-\,^{(W)(m)}J_{AB}=\frac{\sqrt{|\eta|}}{2}\varepsilon_{ABCDE}\frac{1}{4}M^{[CDE]},
\end{eqnarray} where $M^{[CDE]}=\eta^{CF}\eta^{DG}\eta^{EH}M_{[FGH]}$ and we define $M_{[FGH]}$ as
{\footnotesize{\begin{eqnarray}\label{6.04}\\\nonumber
M_{[FGH]}&=&\left(A_M\,^{(W)}T^M_{[FG}\right)\,_{;\,H]}-2\,^{(W)}T^N_{[FH|}
\,^{(W)}T^M_{N|G]}A_M-2\,^{(W)}T^N_{[GH}\,^{(W)}T^M_{F]N}A_M\\\nonumber
&\,&-\,^{(W)}T^N_{[FH|}\overrightarrow{E}_N(A_{|G]})+\,^{(W)}
T^N_{[GH|}\overrightarrow{E}_N(A_{|F]})+\,^{(W)}T^N_{[FH}\overrightarrow{E}_{G]}(A_{N})-\,^{(W)}T^N_{[GH}\overrightarrow{E}_{F]}(A_{N}).\end{eqnarray}}}
Once we required the gauge condition $^{(lc)}A^N_{;\,N}=0$, it is
preserved in the Weitzenb\"{o}ck representation:
$^{(W)}A^n_{;\,n}=0$
\begin{footnote}{One can show that
$\,^{(lc)}{e^N_n}_{;m}=\,^{(W)}{e^N_n}_{;m}+{e^N_k}K^k_{nm}$, but
since the Weitzenb\"{o}ck connections comply with
$\,^{(W)}{e^N_n}_{;m}=0$, it is possible to express the covariant
derivative of the vielbein in the Levi-Civita representation as a
function of the contortion of Weitzenb\"{o}ck:
$\,^{(lc)}{e^N_n}_{;m}=e^N_kK^k_{nm}$. If we use this on the gauge
condition and we take into account that $A^N=e^N_nA^n$, we can
prove that $^{(lc)}A^N_{;\,N}=0 \Rightarrow
\,^{(lc)}A^n_{;\,n}+K^m_{mn}A^n=\,^{(W)}A^n_{;\,n}=0$. Hence, one
can show the following equalities: {\begin{itemize}\item
$\,^{(W)}{A^n}_{;n}=\,^{(W)}{A^N}_{;N}$, \item
$\bar{e}_K^k\eta^{BC}\,^{(W)}A^K_{;\;BC}=g^{ab}\,^{(W)}A^k_{;bc}$,
\item $e^N_n
\eta^{MC}\,{{\,^{(W)}F_{NM;\,C}}}=g^{mc}\,^{(W)}F_{nm;\,c}$, \item
$\,^{(lc)}\Box A^K=e^K_k \,^{(W)}\Box A^k-e^K_k
g^{bc}\,^{(W)}A^k_{;n}K^n_{bc}$. Furthermore:
$g^{bc} K^n_{bc}=g^{mn}\,
^{(W)}T^c_{cm}$.\end{itemize}}}\end{footnote}.

\section{Gravito-magnetic currents from WIMT}

We intend to explore the flat-spacetime in order to learn
under what conditions we can introduce gravito-magnetic currents
using WIMT on 5D. We shall consider an orthogonal base without
structure and trivial connections.

\subsection{Quantization of the fields}

Starting from the vacuum action (\ref{accionfaraday}) with
$^{(lc)}R=0$, one obtains the dynamics of the
gravito-electromagnetic fields
\begin{eqnarray}\label{dina}
    \Box A^K&=&\eta^{BC} A^K_{;\,BC}=\eta^{BC}
    A^K_{,\,BC}=\\\nonumber
    &=&A^K_{,\,tt}-A^K_{,\,xx}-A^K_{,\,yy}-A^K_{,\,zz}-A^K_{,\,ll}=0,
\end{eqnarray}
where $\overrightarrow{A}=A^K\overrightarrow{E}_K$. The
differential equations (\ref{dina}) are separable, so that we can
propose a solution of the kind
\begin{equation}A^K=T^K(t)X^K(x)Y^K(y)Z^K(z)L^K(l).\end{equation}
The vector field can be written as a Frourier expansion
\begin{eqnarray}\label{expcampo1}\\\nonumber
    \overrightarrow{A}=\int_{K^N_x}\int_{K^N_y}\int_{K^N_z}d^3(K^N)\,e^{-K^N_l\frac{l-l_0}{l_0}}\left[A(\mathbf{K}^N,K^N_l,N)
    e^{i(K^N_x\frac{x}{x_0}+K^N_y\frac{y}{y_0}+K^N_z\frac{z}{z_0}-K^N_t\frac{t}{t_0})}\right.\\\nonumber \left.+B(\mathbf{K}^N,K^N_l,N)
    e^{-i(K^N_x\frac{x}{x_0}+K^N_y\frac{y}{y_0}+K^N_z\frac{z}{z_0}-K^N_t\frac{t}{t_0})}\right]\overrightarrow{E}_N.
\end{eqnarray}
We have expanded the vector field as a function of the components
of the tangent base $\{\overrightarrow{E}_N\}$. These fields
comply with $\Box A^M=\eta^{BC}A^M_{;\,BC}=0$, where the metric
$\eta_{BC}=\underrightarrow{\underrightarrow{\eta}}(\overrightarrow{E}_B,\overrightarrow{E}_C)$
describe an inner product through the application on the elements of
the tangent space. It is clear that the connections are null. The
fields can be expanded with any base with the following
requirement for the polarization vectors:
$\xi_M(\overrightarrow{K}^N,L)\xi^M(\overrightarrow{K}^N,L')=\eta_{LL'}$,
where
$\overrightarrow{\xi}(\overrightarrow{K}^N,M')=\xi^M(\overrightarrow{K}^N,M')\overrightarrow{E}_M$.
In general the choice of the polarization vectors is independent
of  $\overrightarrow{K}^N$, which is the wave vector of the
$N$-component for the field related to the directional propagation
of flat waves.

The expression
$\overrightarrow{\xi}(\overrightarrow{K}^N,M')=\xi^M(\overrightarrow{K}^N,M')\overrightarrow{E}_M$
says in a general manner that the elements
$\xi^M(\overrightarrow{K},M')$ are exactly the vielbein $e^M_{M'}$
that relate the base $\{\overrightarrow{E}_M\}$, with a certain base
$\{\overrightarrow{E}'_{M'}:=\overrightarrow{\xi}(\overrightarrow{K}^N,M')\}$,
with the same normalization. The vector
$\overrightarrow{K}^N:=K^N_t\overrightarrow{E}_t+K^N_x\overrightarrow{E}_x+K^N_y\overrightarrow{E}_y+K^N_z\overrightarrow{E}_z+iK^N_l\overrightarrow{E}_l$,
is a field in a 5D vacuum and complies with $|K^N|^2=0$, 
which describes the propagation of a light cone. Is usual to propose the
radiation gauge $A^0=0$ and $A^N_{;\,N}=0$. After taking into
account the isotropy of the space it is obvious that (for
$N,M=1,2,3$): $K^N_i=K^M_j$, where $i,j=1,2,3$.

The field vector $A(\mathbf{K}^N,K^N_l,N)$ depends on the
component of the field which we are describing (which we rotulate
with the superscript $N$), with the spatial part of the wave
vector $\mathbf{K}^N$ and with the extra dimensional component of
$K^N$: $K^N_l$. An important point is the second quantization of
the field, from which is guaranteed the interpretation of this
field as an hermitian operator capable to act on the Fock space
through certain operators of creation and annihilation. Hence, we
shall promote the elements $A(\mathbf{K}^N,K^N_l,N)$ and
$B(\mathbf{K}^N,K^N_l,N)$ to operators which comply with the
condition $\overrightarrow{A}=\overrightarrow{A}^{\dag}$. Hence,
we obtain the relationship
$B^{\dag}(\mathbf{K}^N,K^N_l,N)\overrightarrow{E}_N^{*}=A(\mathbf{K}^N,K^N_l,N)\overrightarrow{E}_N$
\begin{footnote}{The adjoint operator is given by {\scriptsize{\begin{eqnarray}\nonumber
    \overrightarrow{A}^{\dag}=\int_{K^N_x}\int_{K^N_y}\int_{K^N_z}d^3(K^N)\,e^{-K^N_l\frac{l-l_0}{l_0}}\left[A^{\dag}(\mathbf{K}^N,K^N_l,N)
    e^{-i(K^N_x\frac{x}{x_0}+K^N_y\frac{y}{y_0}+K^N_z\frac{z}{z_0}-K^N_t\frac{t}{t_0})}\right.\\\nonumber
    \left.+B^{\dag}(\mathbf{K}^N,K^N_l,N)e^{i(K^N_x\frac{x}{x_0}+K^N_y\frac{y}{y_0}+K^N_z\frac{z}{z_0}-K^N_t\frac{t}{t_0})}\right]\overrightarrow{E}_N^{*},
\end{eqnarray}}} so that we see from the expression (\ref{expcampo1}), that $\overrightarrow{A}=\overrightarrow{A}^{\dag}$ it is fulfilled when
{\scriptsize{\begin{enumerate}\item
$A(\mathbf{K}^N,K^N_l,N)\overrightarrow{E}_N=B^{\dag}(\mathbf{K}^N,K^N_l,N)\overrightarrow{E}_N^{*}$,
\item
$B(\mathbf{K}^N,K^N_l,N)\overrightarrow{E}_N=A^{\dag}(\mathbf{K}^N,K^N_l,N)\overrightarrow{E}_N^{*}$.\end{enumerate}}}
We see that both conditions are equivalent.\\ \\
With respect to the interpretation of $\overrightarrow{E}_N^{*}$,
we claim that it is $\overrightarrow{E}_N^{*}
=(e^n_N\overrightarrow{e}_n)^{*}=e^{n\,*}_N\,\overrightarrow{e}_n=e^n_N\overrightarrow{e}_n$.}\end{footnote},
where in our case $\overrightarrow{E}_N^{*}$ is merely symbolic
because the bases are given by real vector fields. Hence
\begin{eqnarray}\label{expcampo2}\\\nonumber
    \overrightarrow{A}=\int_{K^N_x}\int_{K^N_y}\int_{K^N_z}d^3(K^N)\,e^{-K^N_l\frac{l-l_0}{l_0}}\left[A(K^N)e^{i(K^N_x\frac{x}{x_0}
    +K^N_y\frac{y}{y_0}+K^N_z\frac{z}{z_0}-K^N_t\frac{t}{t_0})}\overrightarrow{E}_N\right.\\\nonumber \left.+A^{\dag}(K^N)
    e^{-i(K^N_x\frac{x}{x_0}+K^N_y\frac{y}{y_0}+K^N_z\frac{z}{z_0}-K^N_t\frac{t}{t_0})}\overrightarrow{E}_N^{*}\right].
\end{eqnarray}
Using (\ref{expcampo2}) we obtain the canonical momentum
$\Pi^N:=\frac{\delta\mathcal{L}}{\delta(A_{N,\,0})}$
{\footnotesize{\begin{eqnarray}&\,&\\\nonumber\Pi^N&=&\eta^{N(N)}A_{(N),\,0}={A^N}_{,\,0}\\\nonumber
&=& i
\int_{K^N_x}\int_{K^N_y}\int_{K^N_z}d^3(K^N)\,e^{-K^N_l\frac{l-l_0}{l_0}}\left[\frac{-K^N_t}{t_0}A(\mathbf{K}^N,K^N_l,N)
e^{i(K^N_x\frac{x}{x_0}+K^N_y\frac{y}{y_0}+K^N_z\frac{z}{z_0}-K^N_t\frac{t}{t_0})}\overrightarrow{E}_N\right.\\\nonumber
&\,&
\,\,\,\,\,\,\,\,\,\,\,\,\,\,\,\,\,\,\,\,\,\,\,\,\,\,\,\,\,\,\,\,\,\,\,\,\,\,\,\,\,\,\,\,\,\,\,\,\,\,\,\,\,\,\,\,\,\,\,\,\,\,
\,\,\,\,\,\,\,\,\,\,\,\,\,\,\,\,\,\,\left.+\frac{K^N_t}{t_0}A^{\dag}(\mathbf{K}^N,K^N_l,N)e^{-i(K^N_x\frac{x}{x_0}+K^N_y\frac{y}{y_0}
+K^N_z\frac{z}{z_0}-K^N_t\frac{t}{t_0})}\overrightarrow{E}_N^{*}\right],\end{eqnarray}}}
which complies with the algebra\\
$[A^N(t,\mathbf{x},l),\Pi^M(t,\mathbf{x}',l)]=\,i\,\,a\,b\,(2\pi)^3\,\eta^{NM}\,\delta(\mathbf{x}-\mathbf{x}')\,e^{-K_l\frac{l-l_0}{l_0}}$,\\
$[A^N(t,\mathbf{x},l),A^M(t,\mathbf{x}',l)]=[\Pi^N(t,\mathbf{x},l),\Pi^M(t,\mathbf{x}',l)]=0$,\\
$[A(\mathbf{K}^N,K^N_l,N),A^{\dag}(\mathbf{K}'^M,K'^M_l,M)]=a\,\delta^{NM}\,\delta(\mathbf{K}^N-\mathbf{K}'^M)\,\delta(K^N_l-K'^M_l)$,\\
$[A(\mathbf{K}^N,K^N_l,N),A(\mathbf{K}^M,K^M_l,M)]=[A^{\dag}(\mathbf{K}^N,K^N_l,N),A^{\dag}(\mathbf{K}^M,K^M_l,M)]=0$,\\
$T_{K^N}(t)\,T^{*}_{K^N}(t)_{,\,t}-T^{*}_{K^N}(t)\,T_{K^N}(t)_{,\,t}=i\,b$.\\
\\We have used the notation $\mathbf{x}=(x,y,z)$, $A^N(t,\mathbf{x},l)$
and $\Pi^M(t,\mathbf{x}',l)$ for the components of the
corresponding fields. Furthermore, $A(\mathbf{K}^N,K^N_l,N)$ is an
annihilation operator  and $A^{\dag}(\mathbf{K}^N,K^N_l,N)$ is a
creation operator. For the constant values
$b=2\,\frac{K^N_t}{t_0}$ and $a=\frac{t_0}{2\,(2\pi)^3\,K^N_t}$,
we obtain{\footnotesize{\begin{eqnarray}\label{6.05}\\\nonumber
[A^N(t,\mathbf{x},l),\Pi^M(t,\mathbf{x}',l)]&=&i\,\eta^{NM}\,\delta(\mathbf{x}-\mathbf{x}')\,e^{-K^N_l\frac{l-l_0}{l_0}},\\\nonumber
[A(\mathbf{K}^N,K^N_l,N),A^{\dag}(\mathbf{K}'^M,K'^M_l,M)]&=&\delta^{NM}\frac{t_0}{2\,(2\pi)^3\,K^N_t}\,\delta(\mathbf{K}^N-\mathbf{K}'^M)\delta(K^N_l-K'^M_l).
\end{eqnarray}}}
These expressions extend to the 5D Minkowski spacetime the
canonical quantization obtained in a 4D
spacetime\begin{footnote}{In order to avoid any problems of the
commutators, we shall use the quantization of Gupta \cite{gupta}
and Bleuler \cite{bleuler}: $\left<
A^N_{;\,N}\right>=0.$}\end{footnote}. Therefore, we can transform
the commutators in the Fock space as second rank tensors, in the
following manner:
\begin{eqnarray}\label{6.06}\nonumber
[A^n(t,\mathbf{x},l),\Pi^m(t,\mathbf{x}',l)]&=&i\,g^{nm}\,\delta(\mathbf{x}-\mathbf{x}')\,e^{-K^n_l\frac{l-l_0}{l_0}},\\\nonumber
[A(\mathbf{K}^n,K^n_l,n),A^{\dag}(\mathbf{K}'^m,K'^m_l,m)]&=&g^{nm}\frac{t_0}{2\,(2\pi)^3\,K^n_t}\,\delta(\mathbf{K}^n-\mathbf{K}'^m)\delta(K^n_l-K'^m_l),
\end{eqnarray}
where $\Pi^m=\bar{e}^m_M\Pi^M$ and $K^m=\bar{e}^m_MK^M$.
These expressions provide us with the algebra in an arbitrary metric
obtained from a 5D Minkowski spacetime, which is free of
structure. In order to illustrate the formalism, in the following
section we shall apply it to a de Sitter expansion, which
describes the early inflationary universe.

\section{An example: monopoles from a 5D Minkowski spacetime with contortion}

We shall study an example in which we take as an initial spacetime
a 5D Minkowski described with the cartesian coordinates
$\phi(p)=(t,x,y,z,l)_p$, in which the tensor metric is given by
the orthonormal matrix $\eta^{AB}$ (\ref{metri}). We choose a IB
which is not coordinated, and the structure coefficients are given
by $C^I_{I0}=\frac{-e^{-N}}{l^2}$, $C^I_{I4}=\frac{-e^{-N}}{l^2}$
and $C^0_{04}=\frac{-1}{l^2}$. The vielbein transforms as
$\bar{e}^n_N=diag(1/l,e^{-N}/l,e^{-N}/l,e^{-N}/l,1)$, so that we
obtain the metric
\begin{eqnarray}[g]_{ab}=\left(%
\begin{array}{ccccc}
  \psi^2(l) & 0 & 0 & 0 & 0 \\
  0 & \psi^2(l)\,e^{2N} & 0 & 0 & 0 \\
  0 & 0 & \psi^2(l)\,e^{2N} & 0 & 0 \\
  0 & 0 & 0 & \psi^2(l)\,e^{2N} & 0 \\
  0 & 0 & 0 & 0 & -1 \\
\end{array}%
\right),\end{eqnarray} with $\psi^2(l)=l^2/l_0^2$. The final
spacetime is described by a coordinated base. This implies that
the Weitzenb\"{o}ck torsion in the final spacetime will be
nonzero. This torsion will be a possible geometrical source for
the emergence of gravito-magnetic monopoles, once it has been made
a constant foliation on the extra noncompact coordinate. The
current components in this case are given by
\begin{eqnarray}\label{pichi}^{(lc)(m)}J_i&=&C^j_{0i,\,k}A_j \left(1-\delta^k_i\right)=0, \label{lc} \\
^{(W)(m)}J_i&=&\epsilon_{ijk}
\partial_j A_k \,e^{-H t},\end{eqnarray}
and $^{(W)(m)}J_0=^{(lc)(m)}J_0=0$. Notice that the spatial
components of the magnetic currents decay with time in the
Weitzenb\"ock representation. The study of the dynamics for the
field fluctuations during a de Sitter inflation in the framework
of Gravito-electromagnetic Inflation (GEMI) was studied with detail in \cite{agustinmembiela}, and
goes beyond the scope of this paper. However, as has been
demonstrated in eq. (\ref{lc}), from the point of view of the
Levi-Civita representation there are no currents related to
gravito-magnetic sources.

\section{Conclusions}

We have extended the WIMT formalism to GEMI with the aim to show
that gravito-magnetic currents may be obtained, at least in a
Weitzenb\"ock representation. The WIMT formalism was introduced
with the idea to generalize the foliation method in the Induced
Matter Theory of gravity, in which foliations which are not static as a
result are very difficult to deal with. With the WIMT formalism, one can make
static foliations from a 5D curved spacetime on which one defines
a 5D vacuum from the point of view of a Weitzenb\"ock
representation. Once done the foliation is possible to pass to the
representation of Levi-Civita. It opens a huge versatility to make
static foliations to obtain arbitrary 4D hypersurfaces from 5D
curved manifolds, which could be very important for quantum field
theories, gravitation, cosmology, etc.

In particular, we have centered our study of the WIMT in the dual
formalization of the GEMI applied to the cosmology of the early
inflationary universe. We have obtained nonzero gravito-magnetic
currents with the representation of Weitzenb\"ock. The currents
decrease exponentially with time with the expansion of the
universe, so that at the end of inflation they become negligible. This
should be agree with present observations. However, these currents
are null in a Levi-Civita representation because in this
geometrical representation the coordinate base has no structure
or torsion. In a future work we shall study an example where
gravito-magnetic sources are nonzero in any
representation\cite{ruso2}.

\section*{Acknowledgements}

\noindent J. M. Romero and M. Bellini acknowledge CONICET
(Argentina) and UNMdP for financial support.

\end{document}